\documentclass[prd,twocolumn,showpacs,preprintnumbers,amsmath,amssymb,superscriptaddress,floatfix,nofootinbib]{revtex4}

\usepackage{graphicx}
\usepackage{subcaption}
\usepackage{epstopdf}
\usepackage{bm}
\usepackage{amsmath}
\usepackage{amsfonts}
\usepackage{amssymb}
\usepackage{color}
\usepackage{multirow}
\usepackage{footnote}
\usepackage{siunitx}
\usepackage{newunicodechar}

\usepackage[colorlinks, citecolor=blue,anchorcolor=red,menucolor=red,linkcolor=red,filecolor=red,runcolor=red,urlcolor=blue,frenchlinks=red]{hyperref}
\begin{document}
\title{The Study of Pole Trajectory within a bare state in the coupled channel model}
\author{Wei Hao}
\email{haowei@nankai.edu.cn}
\affiliation{School of Physics, Nankai University, Tianjin 300071, China}

\author{Jia-Jun Wu}
\email{wujiajun@ucas.ac.cn}
\affiliation{School of Physical Sciences, University of Chinese Academy of Sciences, Beijing 100049, China}
\affiliation{Southern Center for Nuclear-Science Theory (SCNT), Institute of Modern Physics, Chinese Academyof Sciences, Huizhou 516000, China}

\author{Jin-Lin Fu}
\email{jinlin.fu@ucas.ac.cn}
\affiliation{School of Physical Sciences, University of Chinese Academy of Sciences, Beijing 100049, China}
%\date{June}

\begin{abstract}

We investigate two-particle scattering and two-particle scattering with a bare basis state using Hamiltonian Effective Field Theory (HEFT).  
We analyze the distribution of two-body scattering poles in the momentum and energy planes under relativistic conditions. 
Compared to the non-relativistic case, there are significant differences in the distribution of bound state poles and resonance poles in the relativistic case, primarily due to the square root term in the relativistic formula.  
By considering pure two-particle scattering, we examine the relationship between the form factor and the number of poles. 
Additionally, we clearly elucidate the effects of attractive and repulsive interactions on the bound state poles and resonance poles.  
More importantly, we extend our model by including a bare state and explore the poles originating from the bare state or coupled channels through the trajectories of pole positions, as well as the compositeness of bound states.

\end{abstract}

\maketitle

\section{Introduction}

Understanding the nonperturbative aspects of Quantum Chromodynamics (QCD), such as confinement and chiral symmetry, relies heavily on the investigation of hadron resonance spectra and decay widths.
Bound states and resonances are fundamental to the study of hadron states. 
These states often couple strongly with continuum states, complicating the analysis of such systems. 
Extracting resonance parameters from reaction data remains a critical task in hadron physics~\cite{Suzuki:2008rp}. 
Techniques such as dispersion relations, $K$-matrix methods, and dynamical models are employed to analytically continue partial-wave amplitudes into the complex energy plane, enabling precise determination of resonance parameters~\cite{Suzuki:2008rp}. 
The identification of resonance poles and their residues is essential for these analyses. 
Determining pole positions plays a pivotal role in both theoretical and experimental studies, as it helps distinguish between bound states, virtual states, and resonances. 
Moreover, understanding pole structures is crucial for elucidating the properties of certain atomic nuclei~\cite{Hammer:2019poc, Hammer:2017tjm}. 
The scattering $T$-matrix serves as a key tool for accurately determining pole positions. 
A significant challenge in scattering theory is understanding the influence of poles on multiple Riemann sheets on observable scattering phenomena. 
Numerous studies have addressed this issue, providing valuable insights~\cite{Pearce:1988rk,Hyodo:2013iga,Doring:2009yv,Guo:2009ct,Li:2022aru,Ikeda:2011dx,Chen:2023eri,Doring:2009yv,Wu:2017qve,Dietz:2023kbs,Abell:2021awi,He:2005ey,Lage:2009zv,Bernard:2010fp,Guo:2012hv,Li:2012bi,Nieves:2001wt}.

\begin{figure}
    \centering
    \begin{subfigure}{0.2\textwidth}
        \centering
        \includegraphics[width=\textwidth]{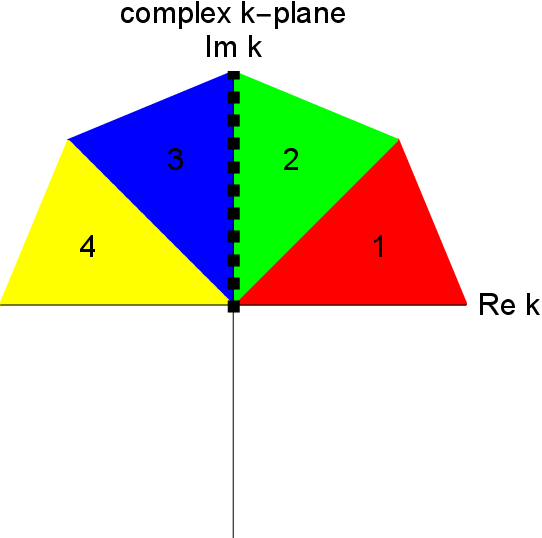}
        \caption{}%{Subfigure A}
        \label{fig:a1}
    \end{subfigure}
    \begin{subfigure}{0.2\textwidth}
        \centering
        \includegraphics[width=\textwidth]{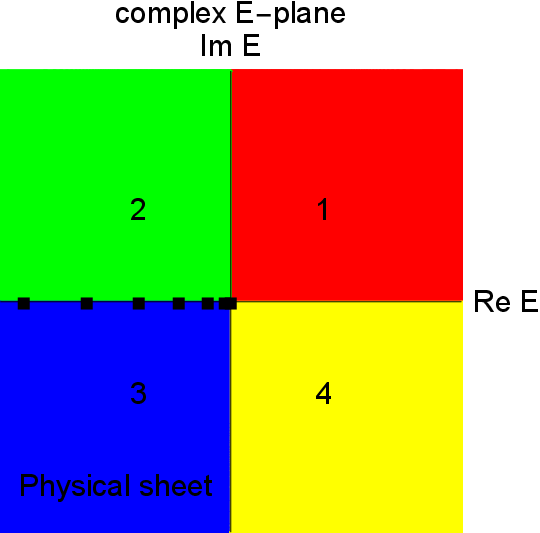}
        \caption{}%{Subfigure B}
        \label{fig:b1}
    \end{subfigure}
    
    \begin{subfigure}{0.2\textwidth}
        \centering
        \includegraphics[width=\textwidth]{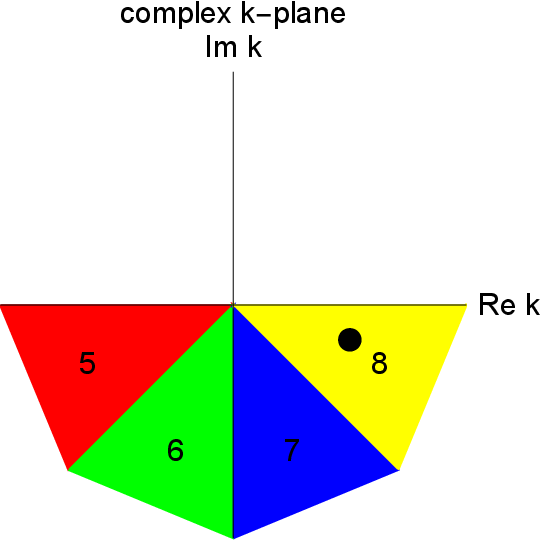}
        \caption{}%{Subfigure C}
        \label{fig:c1}
    \end{subfigure}
    \begin{subfigure}{0.2\textwidth}
        \centering
        \includegraphics[width=\textwidth]{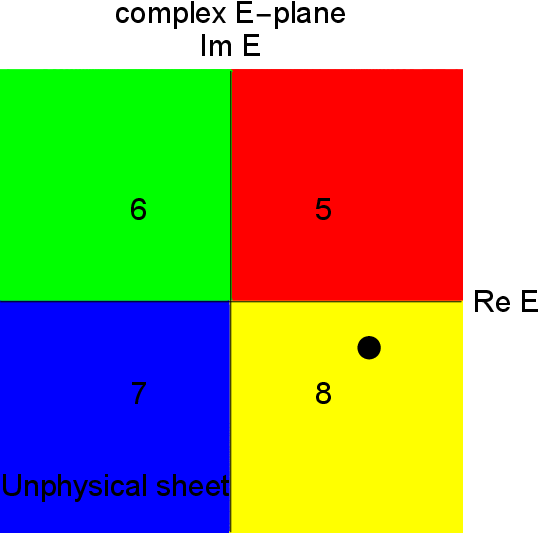}
        \caption{}%{Subfigure D}
        \label{fig:d1}
    \end{subfigure}
    
\caption{The complex momentum $k$-plane and its corresponding complex energy $E$-plane in non-relativistic two-body scattering. The solid squares (circles) represent the bound state (resonance) poles.}
\label{fig:nr-rs}
\end{figure}

As we know, the properties of the non-relativistic two-body scattering and multi-body scattering is very well studied. 
In scattering theory, the scattering $T$-matrix can be defined on complex momentum $k$-plane and energy $E$-plane. 
Due to the double-valued properties of the quadratic relation between $k$ and $E$, i.e., $k=|2mE|^{1/2}e^{i\phi_E/2}$, the scattering $T$-matrix will be a single-valued function in momentum $k$ but double-valued function in energy $E$ which can be seen in Fig.~\ref{fig:nr-rs}. 
The complex $E$-plane can be divided into two Riemann sheets. 
The physical sheet (first sheet) is defined by the range of the phase $0\leq\phi_E\leq2\pi$ and the unphysical sheet (second sheet) by $2\pi\leq\phi_E\leq4\pi$.

On the different Riemann sheet, one can find singularities of the $S$-matrix which correspond to bound state, resonance and virtual state.
A bound state can be found in the positive imaginary axis of the complex $k$-plane that means the state appear at real axis below threshold on the first Riemann sheet of $E$-plane.
A virtual state is defined by a singularity in the negative imaginary axis of the complex $k$-plane and the corresponding definition is below the threshold on the real axis on the second sheet of the complex $E$-plane. 
On the second Riemann sheet of $E$-plane, a pole on the lower half plane and the real part of the pole above threshold is associated with resonance. 
In general, a resonance has a nonzero imaginary part and has an accompanied pole called a conjugate pole which exists on the upper half of the second Riemann sheet. 
Because resonance is a unstable particle, it can form and decay during the collision.  

\begin{figure}
    \centering
    \begin{subfigure}{0.2\textwidth}
        \centering
        \includegraphics[width=\textwidth]{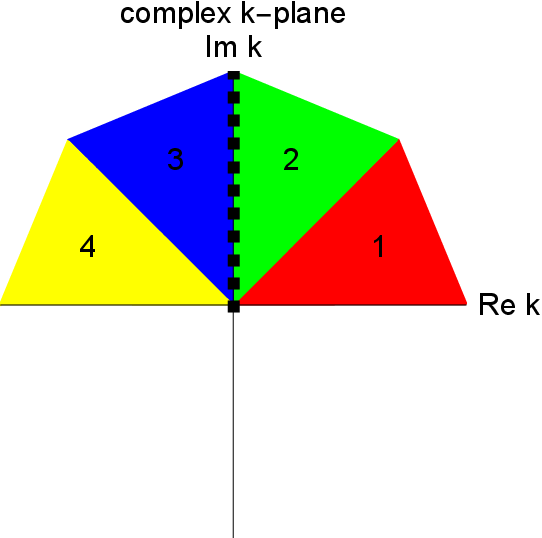}
        \caption{}%{Subfigure A}
        \label{fig:a2}
    \end{subfigure}
    \begin{subfigure}{0.2\textwidth}
        \centering
        \includegraphics[width=\textwidth]{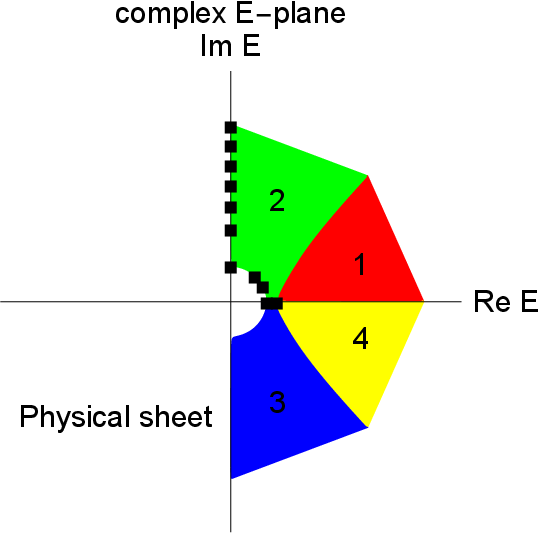}
        \caption{}%{Subfigure B}
        \label{fig:b2}
    \end{subfigure}
    
    \begin{subfigure}{0.2\textwidth}
        \centering
        \includegraphics[width=\textwidth]{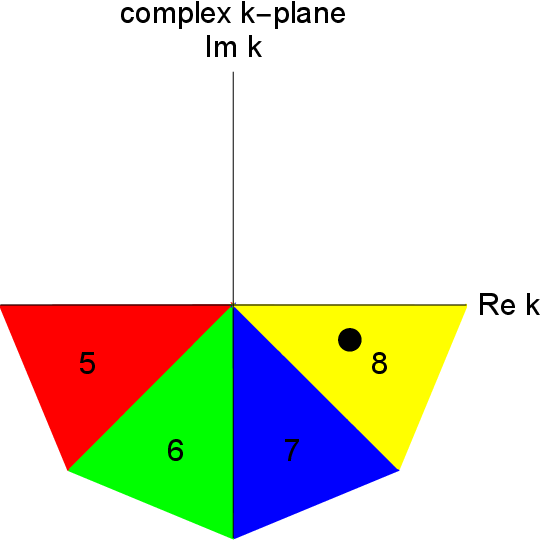}
        \caption{}%{Subfigure C}
        \label{fig:c2}
    \end{subfigure}
    \begin{subfigure}{0.2\textwidth}
        \centering
        \includegraphics[width=\textwidth]{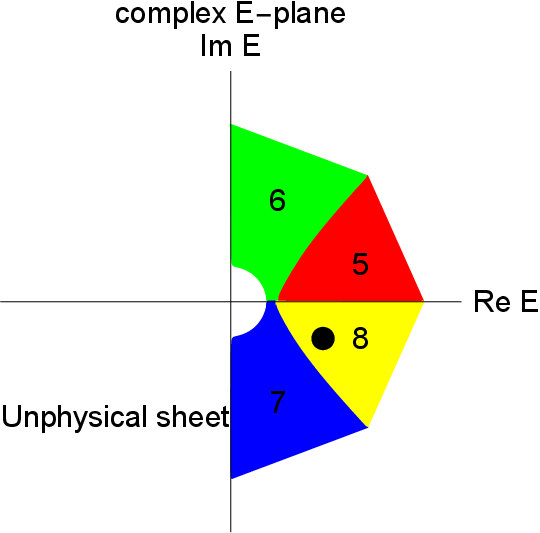}
        \caption{}%{Subfigure D}
        \label{fig:d2}
    \end{subfigure}
    
\caption{The complex momentum $k$-plane and its corresponding complex energy $E$-plane in relativistic two-body scattering. The solid squares (circles) represent the bound state (resonance) poles.}
\label{fig:re-rs}
\end{figure}

In the context of relativistic dynamics, where the energy-momentum relation takes the form $E=\sqrt{m_1^2+k^2}+\sqrt{m_2^2+k^2}$, the properties of bound states and resonances exhibit distinct characteristics compared to the non-relativistic case, as illustrated in Fig.~\ref{fig:re-rs}. 
While bound state poles [solid squares in Fig.~\ref{fig:a2}] remain located on the real axis of the momentum imaginary plane, similar to the non-relativistic scenario, their corresponding positions on the energy plane differ significantly, as shown in Fig.~\ref{fig:b2}.
Furthermore, a comparison of Fig.~\ref{fig:c1}, Fig.~\ref{fig:d1}, Fig.~\ref{fig:c2}, and Fig.~\ref{fig:d2} reveals that although the 7th region of the momentum plane is identical in both non-relativistic (Fig.~\ref{fig:c1}) and relativistic (Fig.~\ref{fig:c2}) cases, the 7th region of the energy plane differs. 
This indicates that the relativistic framework encompasses a broader resonance region than its non-relativistic counterpart. 
Consequently, poles appearing in the 7th region are not purely real, as in bound states, but possess imaginary components, resembling resonances, and may emerge below the threshold. 
While a physical explanation for this phenomenon remains to be fully elucidated, the mathematical structure of the relativistic energy-momentum relation, involving square roots, naturally accommodates such states.

The above analysis indicates that the existence of relativistic kinetic energy terms may lead to the appearance of states with both real and non-zero imaginary parts, similar to resonances, but below the threshold. 
In addition, there are other reasons that can lead to resonances below the threshold. 
People usually believe that if the state below the threshold can further decay, the existence of a resonance at the threshold appears very natural, as the resonance can undergo decay. 
However, in addition to this, there are also some discussions on the existence of resonances under threshold conditions.
For example, in Ref.~\cite{Guo:2009ct}, the authors point out that if the interaction is energy dependent, a state with non-zero imaginary parts can appear below the elastic threshold. 
In Ref.~\cite{Hanhart:2014ssa}, by researching the pole trajectories of resonances, the authors found that the $S$-wave resonances can lead to poles whose real parts below threshold, but imaginary part does not vanish. 
The authors point out that although this is different from the commonly believed phenomenon of resonance states on the real axis below thresholds, it is a real existence. 
Furthermore, in Ref~\cite{Wang:2017agd}, based on the BChPT theory, the authors found that under weak attractive interactions, a ``crazy resonance" with non-zero imaginary parts below threshold will appear.

The Hamiltonian effective field theory (HEFT) is a powerful theoretical tool to study resonance positions, partial decay widths, scattering phase shifts which is related to experimental observation.
It can easily handle scattering problems and incorporate the contribution of bare states into the system. 
On the other hand, HEFT also can build a bridge between finite volume and infinite volume for the multiple channels system, which is equivalent with famous L\"uscher's formula~\cite{Luscher:1985dn,Luscher:1986pf,Luscher:1990ux} upto exponential suppressed correction.  
In the HEFT model, phase shift and inelasticities are derived by solve the LS equation by using the interaction part of the Hamiltonian as the integral equation kernel. 
The pole positions of the state are easily obtained through the T-matrix~\cite{Liu:2016uzk}. 
The spectrum of energy eigenstates can be obtained by solving the Hamiltonian eigen-equation. 
By considering the single bare basis state, the model successfully explained the $\Delta-\pi N$ scattering\cite{Hall:2013qba,Abell:2021awi}, $N^*(1440)-\pi N$ scattering\cite{Wu:2017qve,Liu:2016uzk}, $\Lambda(1405)-\bar{K}N$ scattering\cite{Hall:2014uca}, $N^*(1535)-\pi N$ scattering~\cite{Liu:2015ktc}.

Although multiple articles have conducted a series of research using the HEFT model, people are not particularly clear about the patterns of states that appear in these calculations. 
In Ref.~\cite{Abell:2023qgj}, the authors used a two bare states and one channel system investigating the effects of a second bare basis state in both infinite-volume and finite volume systems to study $\Delta^*$, $N^*(1535)$ and $N^*(1650)$. 
However, there has been a lack of systematic investigation into the role of bare states in generating physical poles (i.e., bound states, resonant states, and virtual states), even in the case of a single bare state coupled to a single channel.
So in this work, we will study the one bare state and one channel system. 
We mainly focus on the study of the variation of poles related to hadron states in the model with the variation of parameters. 
We will systematically discuss the variation patterns of resonances and bound states that appear in the model, as well as their relationship with the parameters. 
We will analyze two modes, the first one named the c-c mode, for which only two-body interactions are considered. 
The second one named the b-cc mode, in this case, not only the two-body interactions but also a bare basis state influence are included.
The interactions between these basis states are parametrized by separable potentials.

The paper is arranged as follows.
The theoretical formalism of the Hamiltonian effective field theory (HEFT) is given in  Section~\ref{sec:formalism}.
We discuss the relationship between the number of pole position and the form factor in Section~\ref{sec:poles}. 
The numerical calculation results and discussions of the pole law, pole trajectory of c-c mode and b-cc mode, and composition distributions are shown in Section~\ref{sec:results}.
In Section~\ref{sec:comp}, we discuss the relationship between  compositeness of the bound states and its binding energy.
In the end, we give the summary  in Section~\ref{sec:summary}.

\section{The brief introduction of coupled-channel formalism}
\label{sec:formalism}

In Hamiltonian Effective Field Theory (HEFT) \cite{Abell:2023qgj,Abell:2023nex,Liu:2016uzk,Wu:2014vma,Liu:2015ktc,Wu:2017qve}, the basis states include two-particle non-interacted states, $|\alpha(\boldsymbol{k})\rangle$ with the three-momentum $\boldsymbol{k}$ in the center-mass frame, and single particle states named bare states, $\left|B\right\rangle$.
Here the interactions between these basis states are parameterised by separable potentials. 
The Hamiltonian of the HEFT framework is constructed as
\begin{equation}
H=H_0+H_I,
\end{equation}
where $H_0$ is the non-interacting Hamiltonian, expressed as
\begin{align}
H_0= & \sum_{B}\left|B\right\rangle m_{B}\left\langle B\right| 
%    \nonumber\\& 
+\sum_\alpha \int d^3 k|\alpha(\boldsymbol{k})\rangle \omega_\alpha(\boldsymbol{k})\langle\alpha(\boldsymbol{k})| ,
\end{align}
and $H_I$ is the interacting part, shown as
\begin{align}
H_{\mathrm{I}}= & \sum_{B} \sum_\alpha \int d^3 k\left\{\left|B\right\rangle G_\alpha^{B}(\boldsymbol{k})\langle\alpha(\boldsymbol{k})|
%\right. \nonumber\\&\left.
%+|\alpha(\boldsymbol{k})\rangle G_\alpha^{B^{*}} \boldsymbol{k})\left\langle B\right|
+h.c.\right\} 
\nonumber\\
& +\sum_{\alpha, \beta} \int d^3 k \int d^3 k^{\prime}|\alpha(\boldsymbol{k})\rangle V_{\alpha \beta}\left(\boldsymbol{k}, \boldsymbol{k}^{\prime}\right)\left\langle\beta\left(\boldsymbol{k}^{\prime}\right)\right| .
\end{align}
Here $G_\alpha^{B}(\boldsymbol{k})$ is for the bare vertex between bare state and two-body system, while $V_{\alpha \beta}\left(\boldsymbol{k}, \boldsymbol{k}^{\prime}\right)$ is for the potential between two-body systems.

By calculating the $T$ matrix based on this Hamiltonian after partial wave, the pole positions related to resonances, bound states or virtual states can be investigated.
It is worthy to mention that here we only consider pure $s$- and $p$-wave interaction, and for simplification we omit the label of the partial wave and reverent factor is absorb into the potential.  
The coupled-channel scattering equations can be expressed as
\begin{align}
& T_{\alpha \beta}\left(k, k^{\prime} ; E\right)=\tilde{V}_{\alpha \beta}\left(k, k^{\prime} ; E\right) \nonumber\\
& \quad+\sum_\gamma \int d q q^2 \frac{\tilde{V}_{\alpha \gamma}(k, q ; E) T_{\gamma \beta}\left(q, k^{\prime} ; E\right)}{E-\omega_\gamma(q)+i \epsilon},
\end{align}
where the $\gamma$ indicates the channel in our model, and $\omega_\gamma(k)=\sqrt{m_{\gamma_M}^2+k^2}+\sqrt{m_{\gamma_N}^2+k^2}$.
$m_{\gamma_M}$ and $m_{\gamma_N}$ are the masses of states $M$ and $N$ in channel $\gamma$. 
It should be point out that, in our calculation, we do not calculate based on specific hadron states, but rather select an arbitrary numerical value to complete the calculation, in order to minimize the dependency of the model and make the calculation more universal. 
So we use $M$ and $N$ to represent the states we considered. 
The coupled-channel potential $\tilde{V}$ is expressed as
\begin{eqnarray}
\tilde{V}_{\alpha \beta}\left(k, k^{\prime} ; E\right)&=&V_{\alpha \beta}\left(k, k^{\prime}\right)+\sum_{B} \frac{G_\alpha^{B \dagger}(k) G_\beta^{B}\left(k^{\prime}\right)}{E-m_{B}},
\end{eqnarray}
where, 
\begin{eqnarray}
G_{MN}^{B}(k)&=&g_{MN}^{B}\frac{5}{\sqrt{\omega_M(k)}}u(k),
\label{eq:balphag-s}
\end{eqnarray}
for $S$-wave and 
\begin{eqnarray}
G_{MN}^{B}(k)&=&\frac{g_{MN}^{B}}{m_M}\frac{k}{\sqrt{\omega_M(k)}}u(k).
\label{eq:balphag-p}
\end{eqnarray}
for $P$-wave interaction, respectively.

The full $T$-matrix can be divided into two parts,
\begin{equation}
T_{\alpha \beta}\left(k, k^{\prime} ; E\right)=t_{\alpha \beta}\left(k, k^{\prime} ; E\right)+T^{bare}_{\alpha \beta}\left(k, k^{\prime} ; E\right).
\end{equation}
The first term is the pure two-body rescattering contribution, $t_{\alpha\beta}(k,k^\prime;E)$, here we named as background $T$-matrix, while the second term include the contribution of bare contribution, $T_{\alpha\beta}^{bare}(k,k^\prime;E)$, named as Resonance $T$-matrix.

\subsection{Background T-matrix}
The $t_{\alpha\beta}(k,k^\prime;E)$ related to the dynamically-generated poles, which is determined by two-body interaction, given by
\begin{align}
& t_{\alpha \beta}\left(k, k^{\prime} ; E\right)=V_{\alpha \beta}\left(k, k^{\prime}\right) \nonumber\\
& \quad+\sum_\gamma \int d q q^2 \frac{V_{\alpha \gamma}(k, q) T_{\gamma \beta}\left(q, k^{\prime} ; E\right)}{E-\omega_\gamma(q)+i \epsilon}.
\end{align}

Here, since we mainly discuss the nature of pole rather than the potential form, we assume that the interaction is parameterized by a separable potential as follows,
\begin{equation}
v_{\alpha \beta}\left(k, k^{\prime}\right)=v_{\alpha\beta}f_\alpha(k)f_\beta(k^\prime).\label{eq:alphaalphav}
\end{equation}
Then the $t$-matrix can be expressed as $t_{\alpha \beta}\left(k, k^{\prime} ; E\right)=f_\alpha(k)\tilde{t}_{\alpha\beta}(E)f_\beta(k^\prime)$, where 
\begin{equation}
\label{form factor-s}
f(k)=\frac{5}{\omega_M(k)}u(k) 
\end{equation}
for the $S$-wave, and 
\begin{equation}
\label{form factor-p}
f(k)=\frac{1}{m_M}\frac{k}{\omega_M(k)}u(k)
\end{equation}
for the $P$-wave interaction, respectively.
The $\tilde{t}$ is solve from the following algebra equation,
\begin{equation}
\tilde{t}_{\alpha \beta}(E)=v_{\alpha \beta}+\sum_\gamma \int d q q^2 \frac{v_{\alpha \gamma} f_\gamma(q)^2 }{E-\omega_\gamma(q)+i \epsilon}\tilde{t}_{\gamma \beta}(E).
\end{equation}
The form factor function $u(k)=(1+\frac{k^2}{\Lambda^2})^{-2}$ is used for removing the ultraviolet divergences in the integration with a regulator parameter $\Lambda$. 
It is a smooth regulator to remove contributions from large momentum\cite{Abell:2023qgj,Abell:2023nex}. 
It is worthy to mention that here on the right side of equal sign, the $\tilde{t}_{\gamma \beta}(E)$ is nothing about the integration, thus it is just a algebra equation, and we can obtain a analytical solution.
In this work, we only conducted qualitative research, so we chose a typical value of $\Lambda=0.8$ GeV.
Then the poles of this background $T$ matrix can be obtained by 
\begin{equation}
\label{pole}
0=\det\left\{
\delta_{\alpha\beta}
-
v_{\alpha \beta}\int d q q^2 \frac{ f^2_\beta(q)}{E-\omega_\beta(q)+i \epsilon}\right\}. 
\end{equation}

\subsection{Resonance T-matrix}

The resonance T-matrix is originated from the bare states, but also related to the background amplitude.
It can be expressed as follows,
\begin{align}
T_{\alpha \beta}^{\mathrm{bare}} & \left(k, k^{\prime} ; E\right) 
%\\& 
=\bar{\mathcal{G}}_\alpha^{B \dagger}(k ; E) A_{BB'}(E) \mathcal{G}_\beta^{B'}\left(k^{\prime} ; E\right),
\end{align}
where $\mathcal{G}_\alpha^{B}(k ; E)$ and $\bar{\mathcal{G}}_\alpha^{B}(k ; E)$ can be recognized as the dress coupling function from bare state to two-body system and and vice versa, respectively,
\begin{align}
\mathcal{G}_\alpha^{B}(k ; E)&=G_\alpha^{B}+\sum_\gamma f_\alpha(k) \tilde{t}_{\alpha \gamma}(E) g_{f, \gamma}^{B}(E),\\
\bar{\mathcal{G}}_\alpha^{B}(k ; E)&=G_\alpha^{B\,*}+\sum_\gamma \bar{g}_{f, \gamma}^{B}(E) \tilde{t}_{\gamma \alpha}(E) f_\alpha(k).
\end{align}
Here, we introduce new variables, $g_{f, \gamma}^{B}(E)$ and $\bar{g}_{f, \gamma}^{B}(E)$, for the simplification as follows, 
\begin{align}
g_{f, \gamma}^{B}(E)&=\int d q q^2 \frac{f_\gamma(q) G_\gamma^{B}(q)}{E-\omega_\gamma(q)+i \epsilon},\\
\bar{g}_{f, \gamma}^{B}(E)&=\int d q q^2 \frac{f_\gamma(q) G_\gamma^{B\,*}(q)}{E-\omega_\gamma(q)+i \epsilon}.
\end{align}
Furthermore, the $A(E)$ is a propagator matrix concerning different bare states, and is expressed as, 
\begin{align}
A^{-1}_{BB'}(E)=\delta_{BB'}\left(E-m_{B}\right)-\bar{\Sigma}_{BB'}(E), 
\end{align}
where $\bar{\Sigma}_{B, B'}(E)$ represents the matrix of all one-loop self-energy interactions between bare states.
It will be expressed as,
\begin{align}
\bar{\Sigma}_{BB'}(E)=\Sigma_{BB'}(E)+\Sigma^I_{BB'}(E).
\end{align}
The $\Sigma_{BB'}(E)$ represents the vertex of the loop just from the interaction between bare states and two-body system, and it reads,
\begin{align}
\Sigma_{BB'}(E)=\sum_\gamma \int d q q^2 \frac{G_\gamma^{B}(q) G_\gamma^{B'\,*}(q)}{E-\omega_\gamma(k)+i \epsilon} ,
\end{align}
In addition, the $\Sigma^I_{B,B'}(E)$ associates with the interactions between two-particle states, which will include the background term.
The explicit form is as follows,
\begin{align}
\Sigma_{BB'}^{\mathrm{I}}(E)=\sum_{\alpha, \beta} \bar{g}_{f, \alpha}^{B}(E) \tilde{t}_{\alpha \beta}(E) g_{f, \beta}^{B'}(E).
\end{align}

In this term, there will be two types of the pole.
One is from the term of $\mathcal{G}_\alpha^{B}(k ; E)$ and $\bar{\mathcal{G}}_\alpha^{B}(k ; E)$, stemming from the pole position of background amplitude, $t(k,k';E)$.
However, we should note that such pole will be exactly cancelled with that in the background amplitude, in other word, once bare state involved, the pole from the background amplitude will be absorbed or shifted~\cite{Li:2022aru,Abell:2023qgj}.
The other pole is from the term $A(E)$, which is the true pole position of the full amplitude, and can be solved from the following equation,
\begin{equation}
\label{pole2}
0=\det\{\delta_{BB'}\left(E-m_{B}\right)-\bar{\Sigma}_{BB'}(E)\}. 
\end{equation}

\subsection{Normalization and probabilities}

For bare state $|B\rangle$ and coupled channel states $|\alpha(k)\rangle$, the eigenvalue equation of the Hamiltonian system can be written as
\begin{align}
H_0 |\alpha(k)>&=(E_{\alpha_M}(k)+E_{\alpha_N}(k)) |\alpha(k)>,\\
H_0 |B>&=m_{B} |B>.
\end{align}

For a bound state $\bar{B}$, the eigenvalue equation of the Hamiltonian is defined as
\begin{align}
H |\bar{B}\rangle&=E_{\bar{B}} |\bar{B}\rangle,\\
|\bar{B}\rangle&=\frac{1}{Z^{1/2}} \left[\sum_{B}c_B|B\rangle+\sum_\alpha\int k^2 dk a_\alpha(k)|\alpha(k)\rangle\right],\label{eq:zzzz}
\end{align}
where $Z$ is normalization constant, $c_B$ and $a_\alpha(k)$ are the bare state and $|\alpha(k)\rangle$ components of the bound state $\bar{B}$, respectively, and $E_{\bar{B}}$ is the mass of the bound state.
In this paper, our aim is to find the nature of pole position related to bare states, thus, to simplify this model, we only consider one bare state and one two-body coupled channel. 
Then the expression of the bound state reads,
\begin{align}
|\bar{B}\rangle&=\frac{1}{Z^{1/2}} \left[|B\rangle+\int k^2 dk a(k)|k\rangle\right],
\label{eq:boundstateZ}
\end{align}
It is clear that the value of $Z$ and $a(k)$ can be determined from the model parameters.
Then the possibilities to have the $B$ in the wave function of the $\bar{B}$ can be calculated by the formula
\begin{equation}
    Z=1-\left.\frac{d\bar{\Sigma}_{BB}(E)}{dE}\right|_{E=E_{\bar{B}}}
    \label{z}.
\end{equation}

\section{The number of pole position vs the form factor}
\label{sec:poles}

In this section, let us first discuss the relationship between the form factor and the pole position.
Typically, the number of the poles of $T$-matrix in the complex plane of energy indeed rely on the form of the form factor through our investigation.

To clarify this point, here we remove the bare state contribution, only focus on the single coupled-channel indicated by $\alpha$.
As shown in Eq.~(\ref{pole}), the pole position of $E$ is satisfied,
\begin{equation}
1=v_{\alpha\alpha}\int dq q^2\frac{f^2_\alpha(q)}{E-\omega_\alpha(q)+i\epsilon}.\label{eq:forpole}
\end{equation}
Obviously, the pole position of $E$ is determined by coupling constant $v_{\alpha\alpha}$ and the form factor $f_\alpha(q)$ in the current model.
As a simple example, we choose a simple form factor as
\begin{equation} 
f_{\alpha}(k)=\frac{2}{(1+(k/\Lambda)^2)^n}.\label{eq:formfactor}
\end{equation}
We can show the number of pole position rely on the value of $n$.
Nevertheless the $\omega_\gamma(q)$ is chosen as non-relativistic form $\omega_\alpha(k)=m_{\alpha_N}+m_{\alpha_N}+\frac{k^2}{2m_{\alpha_N}}+\frac{k^2}{2m_{\alpha_m}}$ or in relativistic form $\omega_\gamma(k)=\sqrt{k^2+m^2_{\alpha_N}}+\sqrt{k^2+m^2_{\alpha_M}}$. 
The calculation results shown that if choosing $n$=1, 2, 3, etc., the number of poles will be 1, 2, 3, etc..
Furthermore, it can be proven that if $n$=1/2, 3/2, 5/2, etc., the corresponding number of poles is 1, 2, 3, etc.. 
Indeed, it can be proven analytically in the non-relativistic case.
For example, when we take $n=1$ in Eq.~(\ref{eq:formfactor}), we can explicitly obtain the integration as follows,
\begin{equation}
\int dq q^2\frac{f^2_\alpha(q)}{E-\omega_\alpha(q)+i\epsilon}=\frac{2\pi i \sqrt{\mu}\sqrt{\frac{k_0^2}{\mu}}+k_0^2-1}{(k_0^2+1)^2}.
\end{equation}
where $\mu=\frac{m_{\alpha_N} m_{\alpha_m}}{\Lambda^2(m_{\alpha_N}+m_{\alpha_M})}$, $k_0=\sqrt{2\mu (E-m_{\alpha_N}-m_{\alpha_M})}$. 
With this expression, we can easily calculate the pole positions, for example, when we assume $m_1=1$ GeV, $m_2=1$ GeV, $\Lambda=1$ GeV and $v=-1$, a bound state pole $E=2-0.597$ GeV will be obtained. 

If we set $n=2$, the integration can also be analytically calculated as 
\begin{eqnarray}
&&\int dq q^2\frac{f^2_\alpha(q)}{E-\omega_\alpha(q)+i\epsilon}\nonumber\\
&=&\frac{\pi \mu (k_0^6+5k_0^4-16i\sqrt{\mu}\sqrt{\frac{k_0^2}{\mu}}+15k_0^2-5)}{4(k_0^2+1)^4}.
\end{eqnarray}
In this case, we will find two pole positions, for example, we take the same value as that in $n=1$ case, the pole positions will be $E=2-0.745-2.105i$ GeV and $E=2-0.058$ GeV. 
Then when we increase the value of $n$, the number of pole position will increase.
It is due to that the power of $k_0$ will increase by enlarge $n$, and the higher powers of polynomial equation usually produce more solutions. 

For the case of relativity, although analytical solutions cannot be obtained, numerical results indicate that similar patterns are still maintained.

It is important to mention that not all of the poles solved from $T$-matrix have physical meanings, since most of them is just from the model itself. 
Usually, people choose poles that are close to the threshold or have imaginary part values within a reasonable range, such as pole value with a width within a few hundred MeV, which are considered meaningful. 
Of course, it requires physical screening of the poles to be more convincing. Based on this consideration, we attempt to screen multiple poles by analyzing the phase shift of physical observations. 

\begin{figure}
    \centering
    \includegraphics[width=0.8\linewidth]{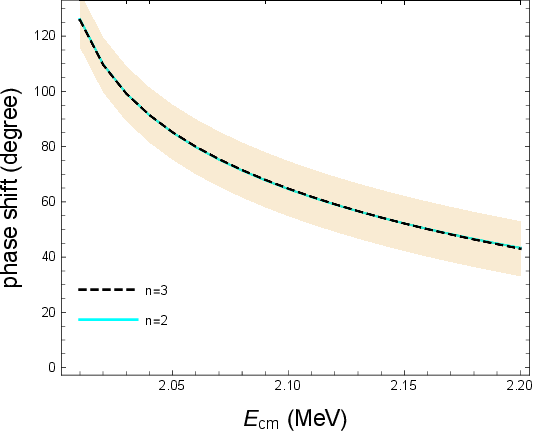}
    \caption{The result of fitting the phase shift of n=3 (black dashed line) in the case of n=2 (cyan solid line). The shaded areas in the figure indicate that there may be a certain range of phase shift.}
    \label{fig:phase1}
\end{figure}

\begin{figure}
    \centering
    \includegraphics[width=0.8\linewidth]{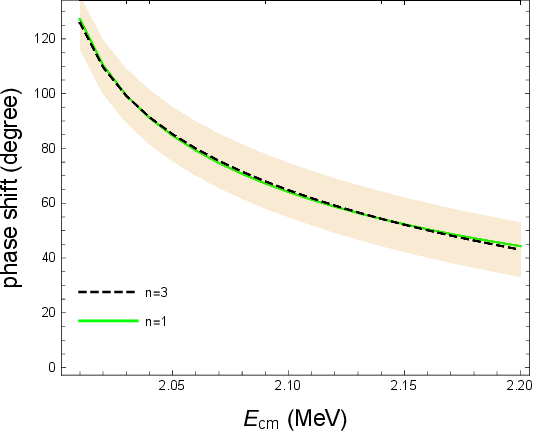}
    \caption{The result of fitting the phase shift of n=3 (black dashed line) in the case of n=1 (green solid line). The shaded areas in the figure indicate that there may be a certain range of phase shift}
    \label{fig:phase2}
\end{figure}

We take a relativistic case as an example.
Our approach is as follows.
Firstly, we choose $n=3$ in Eq.~(\ref{eq:formfactor}) and same values of masses of two particles and $\Lambda$ as before, but using $\omega_\gamma(k)=\sqrt{k^2+m^2_{\alpha_N}}+\sqrt{k^2+m^2_{\alpha_M}}$.
Then three poles will be obtained $1.98433$ GeV, $1.14228-2.36899i$ GeV and $2.22393-1.26388i$ GeV. 
Furthermore, the phase shift trajectory can also be obtained as shown in Fig.~\ref{fig:phase1} (also shown in Fig.~\ref{fig:phase2}) within black dashed line. The shaded areas in the figures indicate that there may be a certain range of phase shift.

Secondly, we use the form factor with $n=2$ and $n=1$ to fit this phase shift, the fitting results are $f_\alpha(k)=\frac{1.99432}{(1.0328+1.6988q^2)^2}$ and $v=-1.1683$, as well as $f_\alpha(k)=\frac{2.008}{(0.40655+1.8901q^2)^1}$ and $v=-0.18356$. 
Meanwhile the phase shifts for the two cases are shown in Fig.~\ref{fig:phase1} with cyan solid line,  and in Fig.~\ref{fig:phase2} with green solid line . 
Then we can obtain the poles in the two cases, there will be two poles $1.98336$ GeV and $1.79308-1.39633i$ GeV in $n=2$ case, while just one pole at $1.97886$ GeV is found in $n=1$ case. 
This indicates that the pole positions controlled by the phase shift in this range is just around $1.98$ GeV, it is a good bound state pole, and other poles are all model dependent, should be neglected.

Through this example, it reminds us that when doing phase shift fitting, we need to pay attention to some poles that may be caused by form factors and may not have physical significance. 
Necessary analysis and screening are important to exclude some poles.
Of course, we only used a very simple form factor as an example for discussion here.
As the form factor becomes more complex, the situation of poles will also become more diverse.

\section{The pole trajectories vs the coupling constants}
\label{sec:results}

In this section, we will discuss the pole trajectories of the coupling constants within the theoretical framework of the section~\ref{sec:formalism}.
It is important to study the pole trajectories for understanding the originate of the poles.
In our model, we have two possible sources to generate the poles.
One is from the bare state, and the other is from the bound state of the two-body system.
It is interesting to show the trajectory of pole from these two type sources.
To make the discussion clearly, we firstly just consider pure one two-body channel without bare state, named as c-c model, while second one includes bare state, named as b-c model.

\subsection{In c-c model}

%In this subsection, we begin analyzing the relationship between poles and interactions within the theoretical framework of the section~\ref{sec:formalism}.
The poles of the c-c model come from the background T-matrix which represent the pure two-body interaction without the contribution of bare state.
As usual, such bound state is recognized as hadronic molecular state, for example, deuteron is the bound state of proton and neutron .
In order to make the conclusion general, we do not choose specific physical states to study the interaction between two bodies, but rather choose two arbitrary masses for $m_M$ and $m_N$. 
We take $m_M=0.2$ GeV and $m_N=0.8$ GeV.
With these masses, the T-matrix poles can be obtained.

Firstly, as discussed in section~\ref{sec:formalism}, our form factors are shown in Eqs.~(\ref{form factor-s}, \ref{form factor-p}) for the $S$-wave and $P$-wave respectively.
This form factor corresponds to the case of $n=2$ as discussed in the previous section, where there will be two pole solutions for each interaction.

\begin{figure}
    \centering
    \includegraphics[width=0.85\linewidth]{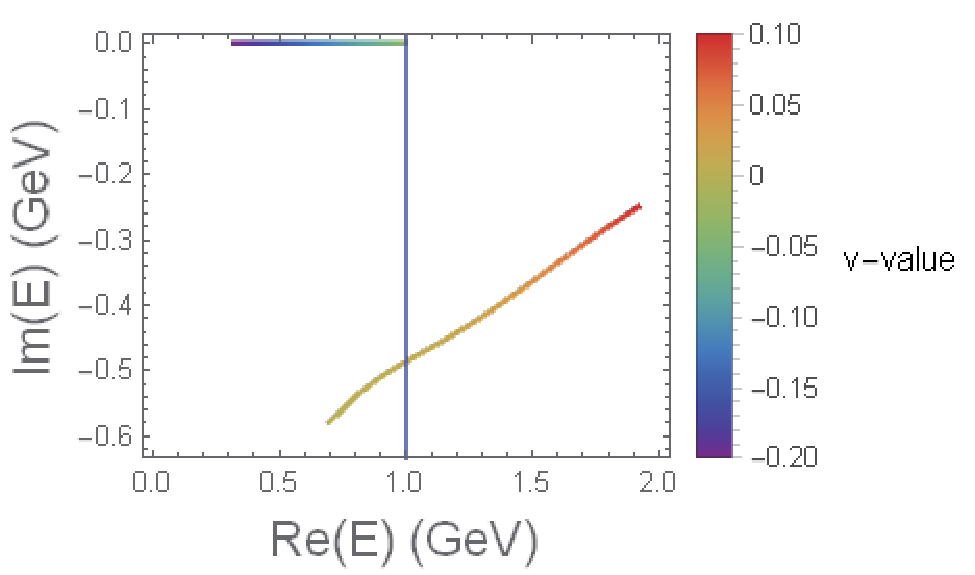}
    \caption{The pole positions with $S$-wave interaction strength.
 The blue line represents the threshold position. Here we set $m_M=0.2$ GeV, $m_N=0.8$ GeV.}
    \label{fig:s-cc}
\end{figure}

Here we provide an example. 
For the $S$-wave interaction, pole trajectories are obtained as shown in Fig.~\ref{fig:s-cc}. 
For the attractive interaction, the bound state poles appears below the threshold (blue solid line).
Furthermore, as the attraction interaction increases, the poles becomes more deeper and be far away from the threshold. 
It is consistent with people's understanding, the more attractive, the deeper bound state. 
When the coefficient of the potential $v_{\alpha\alpha}>-0.0225$, we will find the bound state will disappear and virtual state will be generated in the second Riemann sheet. 
Since in our case, the form factor as shown in Eq.~(\ref{form factor-s}), an additional pole of momentum $q$ is at $im_M$, which will lead to the integration in Eq.~\ref{eq:forpole} infinite at $E=\sqrt{m_N^2-m_M^2}=0.69$ GeV in the second Riemann sheet.
Therefore, the virtual state will always exist when $v_{\alpha\alpha}<0$.
As for the repulsive $S$-wave interaction, i.e., $v_{\alpha\alpha} > 0$, the resonances poles will start from 
below threshold to above threshold as shown in Fig.~\ref{fig:s-cc}.
Finally, it is worth mentioning that the trajectory of this pole is discontinuous at $v_{\alpha\alpha} = 0$. 
This can be easily understood because, to satisfy the condition of Eq.~(\ref{form factor-s}), even if they are on the same Riemann sheet, as $v_{\alpha\alpha}$ changes from $0^-$ to $0^+$, the corresponding integral transitions from $-\infty$ to $+\infty$. 
Consequently, the positions of the poles will not remain the same.

\begin{figure}
    \centering
    \includegraphics[width=0.8\linewidth]{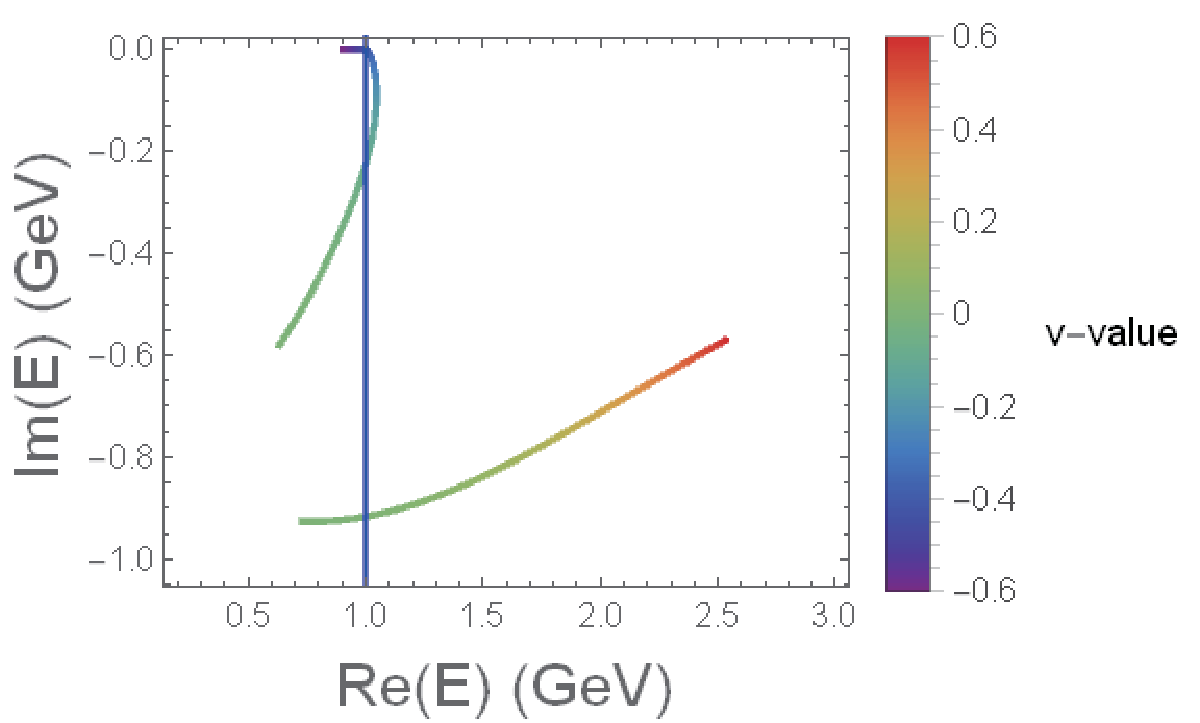}
    \caption{The pole positions with $P$-wave interaction strength.
 The blue line represents the threshold position. Here we set $m_M=0.2$ GeV, $m_N=0.8$ GeV.}
    \label{fig:p-cc}
\end{figure}

For the $P$-wave interaction, the trajectories of pole positions changing with interaction strength are shown in Fig.~\ref{fig:p-cc}. 
It shows that the trajectory of the poles is relatively natural, as the attractive strength increases, the poles gradually shift from the resonance pole to the bound state pole.
As the strength further increases, the bound state will move away from the threshold, become a deeper bound state. 
In this case, we noticed a very strange phenomenon that the resonance state poles appear below the threshold when the interaction strength is relatively small.
As we mentioned in the introduction, this is due to the relativistic effect in the $\omega_\gamma(q)$ changing the possible distribution regions of the solution. 
It corresponds to the pole appearing in the $k$-plane 7th region of Fig.~\ref{fig:c2}. 
When the range of rotation angle $\theta$ is selected in the $k$-plane 7th region, the pole can appear, and this pole corresponds to $E$-plane [Fig.~\ref{fig:d2}], which is the 7th region below the threshold. 
Then leading to the existence of the poles below threshold.

The case of the resonance poles appear below the threshold is clearly not in line with people's usual understanding that the resonances only appear above the threshold without secondary decay.
This issue seems to be less discussed at the meson level, but there are some discussions at the nuclear level~\cite{Lazauskas:2005fy,Marques:2021mqf}. 
In Ref.~\cite{Lazauskas:2005fy}, the authors studied the trajectory of nuclear resonance states under the relativistic interaction model. 
In higher-order Pade approximation, 
the $^3P_0 nn$ resonances trajectory calculated using the the method of analytic continuation in the coupling constant (ACCC) under approximation is very similar to the situation we encountered here.
The paper also shows that when the pade order is small, such as Pade$^{[3,3]}$, the resonance poles represented by the curve will not appear below the threshold. 
But when the pade order is large enough, such as Pade$^{[5,5]}$, the resonance poles represented by the curve will appear below the threshold.

Furthermore, for repulsive $P$-wave interactions, there is no bound states, which is in line with our understanding. 
The motion of the poles indicates that as the repulsive interaction strength increase, the resonance poles will far away from threshold and the imaginary parts of the generated resonances will gradually decrease, similar as that in $S$-wave case.

\subsection{In b-c model}

In this subsection, we concentrate the study on the effects of introducing a bare state to the Hamiltonian. 
Similar to the previous section, in order to make the conclusion general, we did not choose a specific physical mass, but instead used an arbitrary bare state mass value.
We choose the mass of the bare state to be $m_{B_0}=1.3$ GeV which is above the threshold, and $m_{B_0}=0.7$ GeV below the threshold.

As we know, the interaction strength of the $\alpha-\alpha$ and $B-\alpha$ are determined by parameters $v\equiv v_{\alpha\alpha}$ and $g\equiv g^{B}_{\alpha}$ in the model, as shown in Eq.~(\ref{eq:alphaalphav}) and Eqs.~(\ref{eq:balphag-s},\ref{eq:balphag-p}), respectively. 
In order to more clearly distinguish the differences in the effects they bring, we observe by changing the value of $g$ under different values of $v$.

\begin{figure*}
    \centering
    \includegraphics[width=0.9\linewidth]{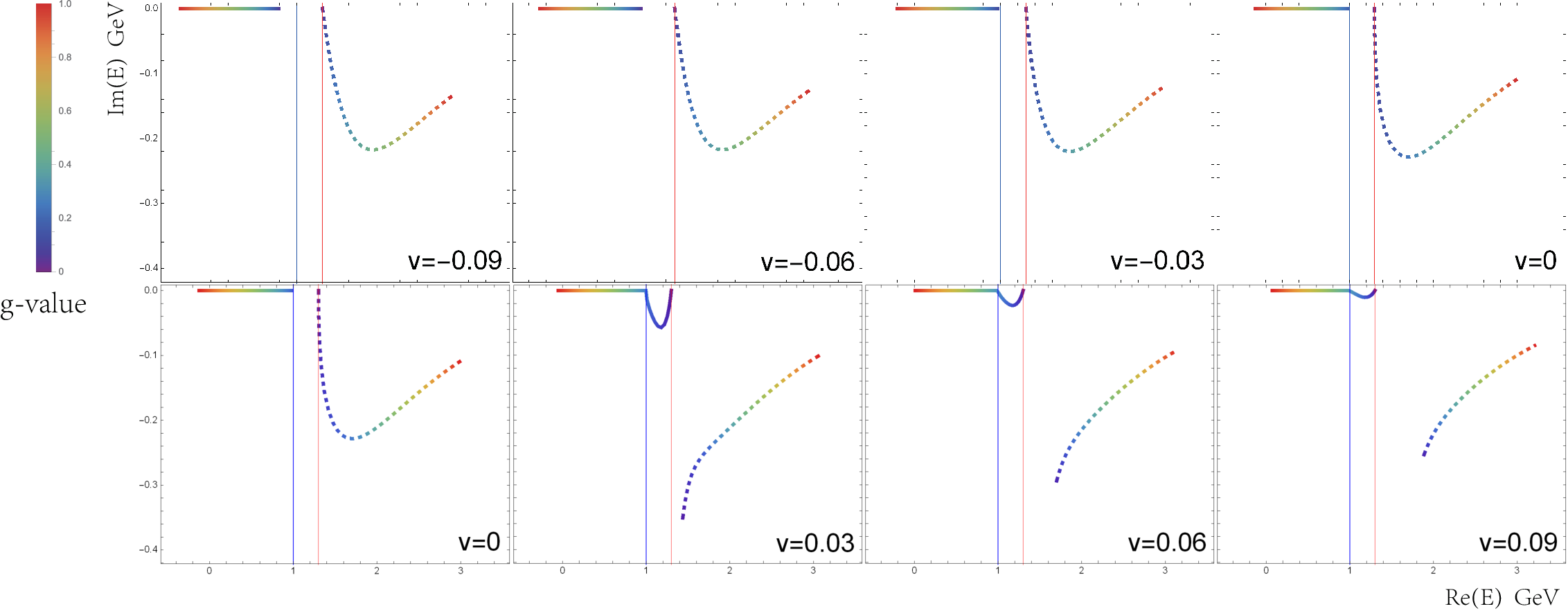}
    \caption{$S$-wave interaction poles as the coupling of the (above threshold)  bare state to $MN$ scattering states. The blue line represents the threshold position, and the red line represents the bare state mass position.}
    \label{fig:s-large}
\end{figure*}

\begin{figure*}
    \centering
    \includegraphics[width=0.9\linewidth]{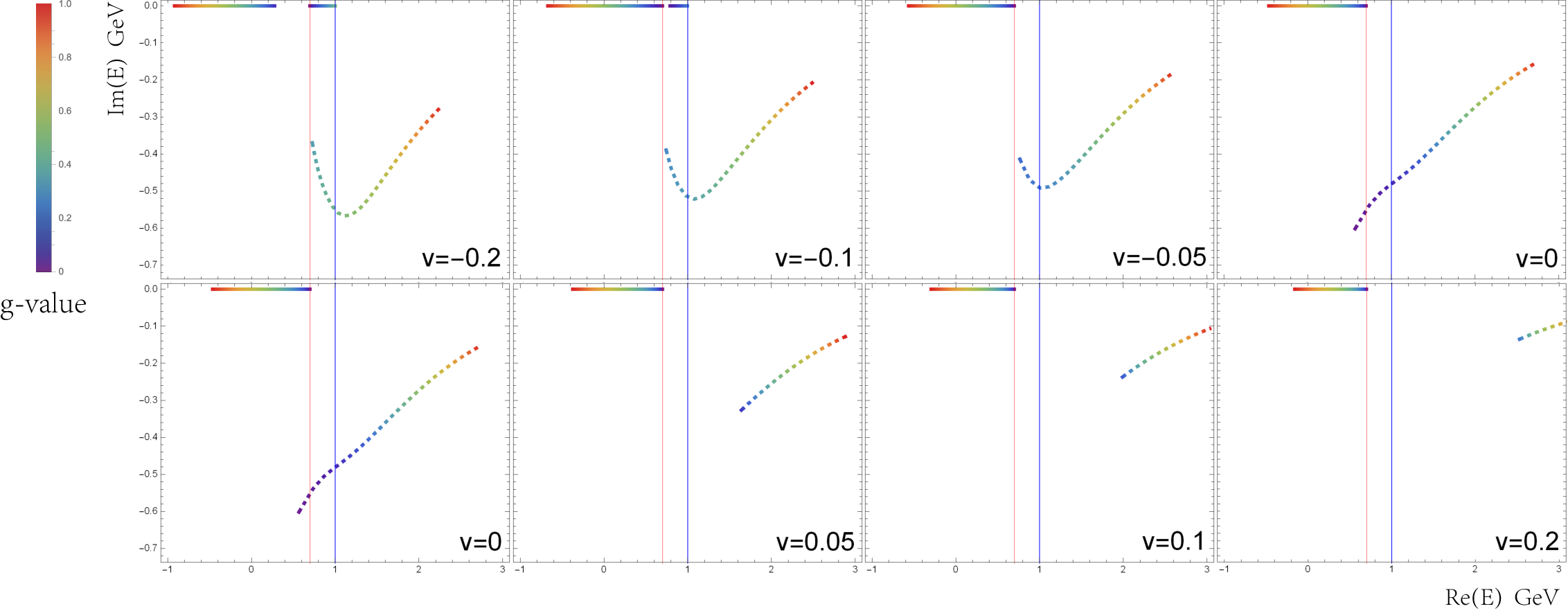}
    \caption{$S$-wave interaction poles as the coupling of the (below threshold)  bare state to $MN$ scattering states. The blue line represents the threshold position, and the red line represents the bare state mass position.}
    \label{fig:s-small}
\end{figure*}

Firstly, we considering the $S$-wave interactions with a bare state above the threshold. 
The $G$ and $f$ function is defined in Eqs.~(\ref{eq:balphag-s}) and (\ref{form factor-s}), respectively.
As shown in Fig.~\ref{fig:s-large}, the pole trajectories is shown clearly. 
We find one pole trajectory always begins with bare state, which is obviously, because the bare state will be a real physical pole once no interaction between the bare state and the coupled channels, i.e., $g=0$.
Thus, we can see that regardless of whether $v$ is attractive or repulsive, a trajectory of singularities always originates from a bare state. 
This also indicates that when a bare state exists, as long as there is an interaction between the bare state and the coupled channel, there will always be a physical singularity closely linked to the bare state.

When $v>0$ shows the repulsive potential, the bound state that forms actually originates from the existence of the bare state, which provides an attractive potential for the two particles. 
This potential can cancel out the repulsive potential brought by $v$, hence forming a bound state. 
Unfortunately, this situation does not yet have a clear physical correspondence as we know. 
We suggest looking for cases where two particles are repulsive and there is a bare state predicted by traditional quark models near the threshold. 
If the attractive potential from the bare state is not strong enough, a resonance state above the threshold will form, which appears between the two vertical red and blue lines in Fig.~\ref{fig:s-large}. 
This has good physical correspondence; many resonance states, such as the $\kappa$ meson and the $\sigma$ meson, are examples of this case. 
Meanwhile, when $v$ is repulsive, the trajectory of singularities along the dotted line is evidently far from the physical region being discussed, which we have found actually originates from the form factor of the model and can thus be ignored.

Next, we consider the case where $v$ is an attractive potential. 
In this case, the trajectory originating from the bare state becomes a resonance as $g$ increases, and a bound state or a virtual state is produced beneath the threshold. 
Here, even if the attractive potential $v$ from the two-body coupled channel is not strong enough, increasing $g$ provides additional attractive potential, allowing a bound state to form. 
In the literature~\cite{Wang:2023ovj}, the explanation for X(3872) reflects such a scenario, and a high-mass $\chi_{c1}(2P)$ state is also predicted.
If the potential of the two-body coupled channel is already sufficient to form a bound state, the bare state will increase the binding energy of this bound state further. 
Actually, $D^*_{s0}(2317)$ can be understood through this framework~\cite{Yang:2021tvc}.
It suggests that part of its deep binding energy over $40$ MeV should come from contributions of the bare state, and we predict that above the bare state mass for this quantum number, there should be a broad resonance state with predominant decay to $DK$.
When the $v$ become a repulsive potential, the bare state will generate a bound state or a resonance for the large and small $g$.
Another resonance state arising from $v$ is located relatively far from the threshold. Given that the model itself is only effective in the vicinity of the threshold, the systematic uncertainty associated with this resonance state is significantly large and can be approximately neglected.

Secondly, let us consider the bare state below the threshold.
As shown in Fig.~\ref{fig:s-small}, the situations of different values of $v$ are rather different.
If the potential $v$ between the coupled channels provides a very strong attractive potential, thereby generating a bound state deeper than the bare state. 
Then since the effective potential generated through the bare state remains attractive below the bare state mass, the bound state produced by $v$ will become even more deeply bound. 
Simultaneously, the mass of the physical state corresponding to the bare state will be pushed higher, potentially turning into a virtual state or a resonance. 
This is precisely the result demonstrated in Fig.~\ref{fig:s-small} for $v = -0.2$. 
When the attractive potential of $v$ decreases, i.e., $v = -0.1$, the bound state produced by $v$ lies above the bare state mass. 
In this case, the effective potential involving the bare state becomes repulsive above the bare state mass, causing the bound state generated by $v$ to become shallower, potentially turning into a virtual state or a resonance. 
If $v$ further decreases and even becomes a repulsive potential, the bare state continues to exist as a bound state, while the other state may transform into a virtual state or a resonance.
Such virtual state or a resonance could be our of the current model, and can be approximately neglected.

\begin{figure*}
    \centering
    \includegraphics[width=1.0\linewidth]{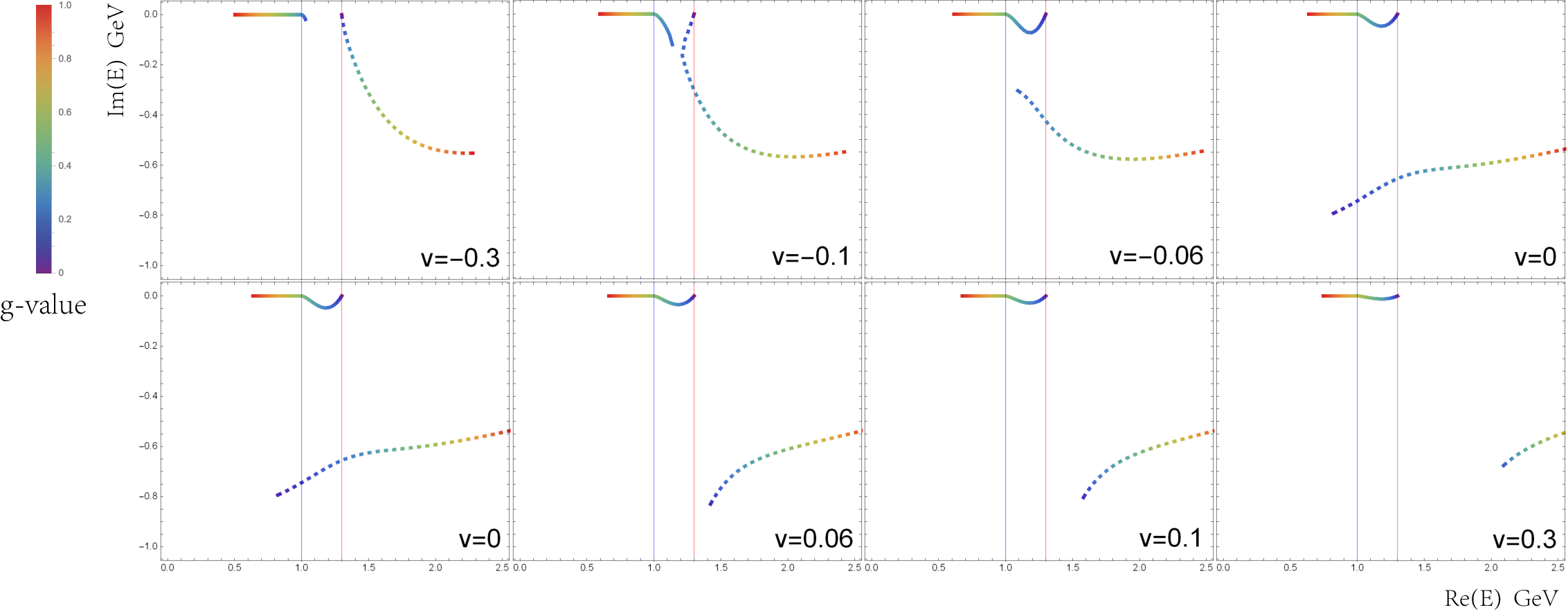}
    \caption{$P$-wave interaction poles as the coupling of the (above threshold)  bare state to $MN$ scattering states. The blue line represents the threshold position, and the red line represents the bare state mass position.}
    \label{fig:p-large}
\end{figure*}

\begin{figure*}
    \centering
    \includegraphics[width=1.0\linewidth]{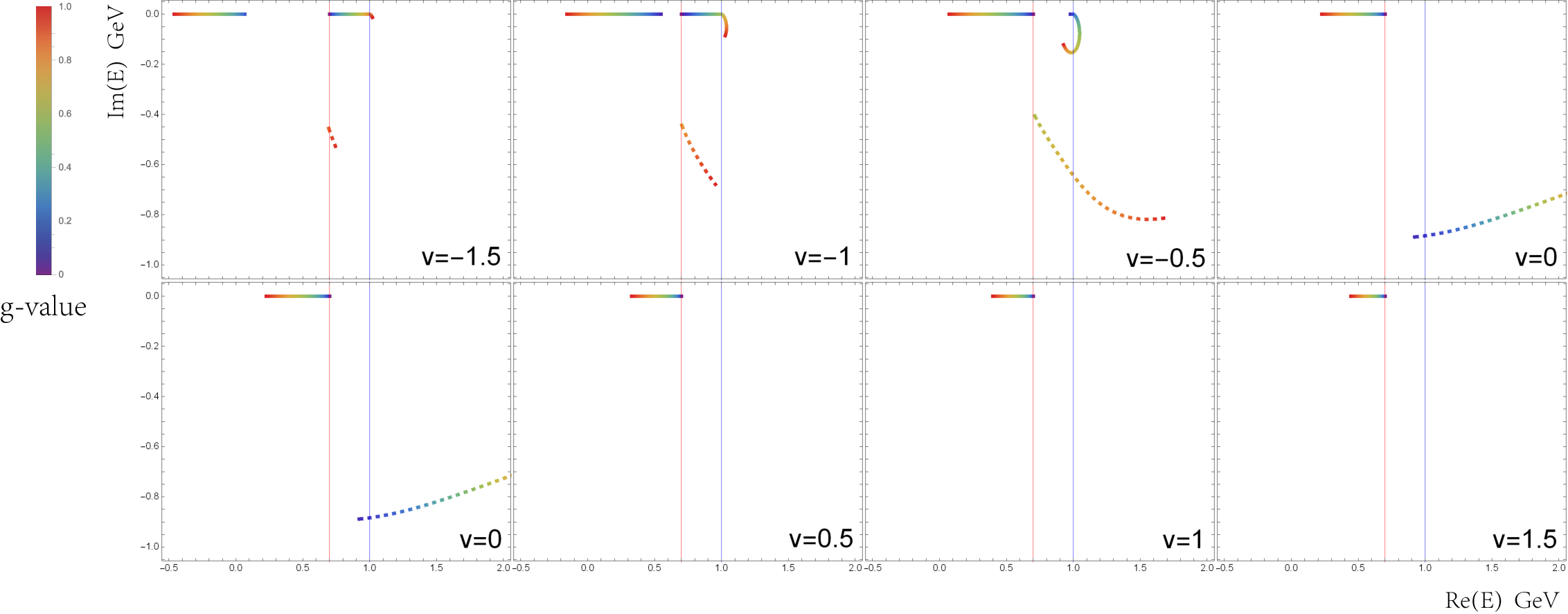}
    \caption{$P$-wave interaction poles as the coupling of the (below threshold)  bare state to $MN$ scattering states. The blue line represents the threshold position, and the red line represents the bare state mass position.}
    \label{fig:p-small}
\end{figure*}

Thirdly, let us discuss the pole in the $P$-wave interaction with a bare state above the threshold.
Now correspondingly, the $G$ and $f$ function is defined in Eqs.~(\ref{eq:balphag-p}) and (\ref{form factor-p}), respectively.
For the repulsive potential case for $v$, the situation is very similar as that in $S$ wave, the bare state will generate a bound state or a resonance dependent on the coupling strength, while another resonance with large width is also in the second Riemann sheet, as shown in the second line of Fig.~\ref{fig:p-large}.
However, for the attractive $v$, when the $v$ is not large enough, the pole trajectories are similar as that for repulsive $v$.
We guess that the $\rho$ meson, the $K^*$ meson and $\Delta(1232)$ baryon, are examples of this case. 
Only when the attractive interaction of $v$ is large enough, the pole trajectory from the bare state will become a resonance state, while the bound state is related to the potential $v$.
Typically, when $v$ only can generate a virtual pole or a resonance pole as shown in Fig.~\ref{fig:p-cc}, two resonances between the threshold of coupled channel and the bare mass will be generated simultaneously as shown $v=-0.1 \sim 0$ in Fig.~\ref{fig:p-large}. 

At last for the bare state below the threshold of coupled channel in $P$-wave, we shown the results in Fig.~\ref{fig:p-small}. 
We will find the bare state will generate a bound state there and there is no additional resonance for $v > 0$.
When $v$ is a attractive integration, the bare state also promise a bound state, but another bound state or a virtual pole or a resonance pole will be generated dependent on the value of $v$ and $g$.

In summary, we have found that if a bare state exists, a physical state will always emerge in its vicinity. 
Depending on the relationship between the bare mass and the threshold, as well as the strength of the interaction, this physical state can manifest as either a bound state or a resonance. 
Additionally, when the attractive interaction between coupled channels is sufficiently strong, another pole may emerge near the threshold.
The properties of this pole are largely determined by the strength of the interaction. 
Notably, due to the fact that the interaction potential contributed by the s-channel of the bare state is repulsive at energies above the bare mass and attractive below it, the relative magnitudes of the bare mass and the threshold directly influence the position and nature of the pole generated by the coupled channels.

\section{The compositeness of the bound states vs the binding energy in $S$ and $P$-wave}
\label{sec:comp}

In this section, let us discuss the compositeness of the bound state generated including the bare state.
As discussed in the previous, we find the bound state could be mainly from the attractive potential between coupled channels or just because of the bare state.
Thus, it is interesting to investigate the $1/Z$ of such bound state.
We know that the compositeness of a state can be determined by analyzing the wave functions of each component of the bound state. 
Here, we calculated the relationship between the bound state components, $1/Z$, (i.e., one minus the compositeness) and the binding energy for both $S$-wave and $P$-wave cases. 
Specifically, it means the possibility of the bare state $B$ component in the dressed state $\bar{B}$, i.e., the value of $1/Z$ defined in Eq.~(\ref{eq:boundstateZ}). 

\begin{figure*}
    \centering
    \includegraphics[width=0.8\linewidth]{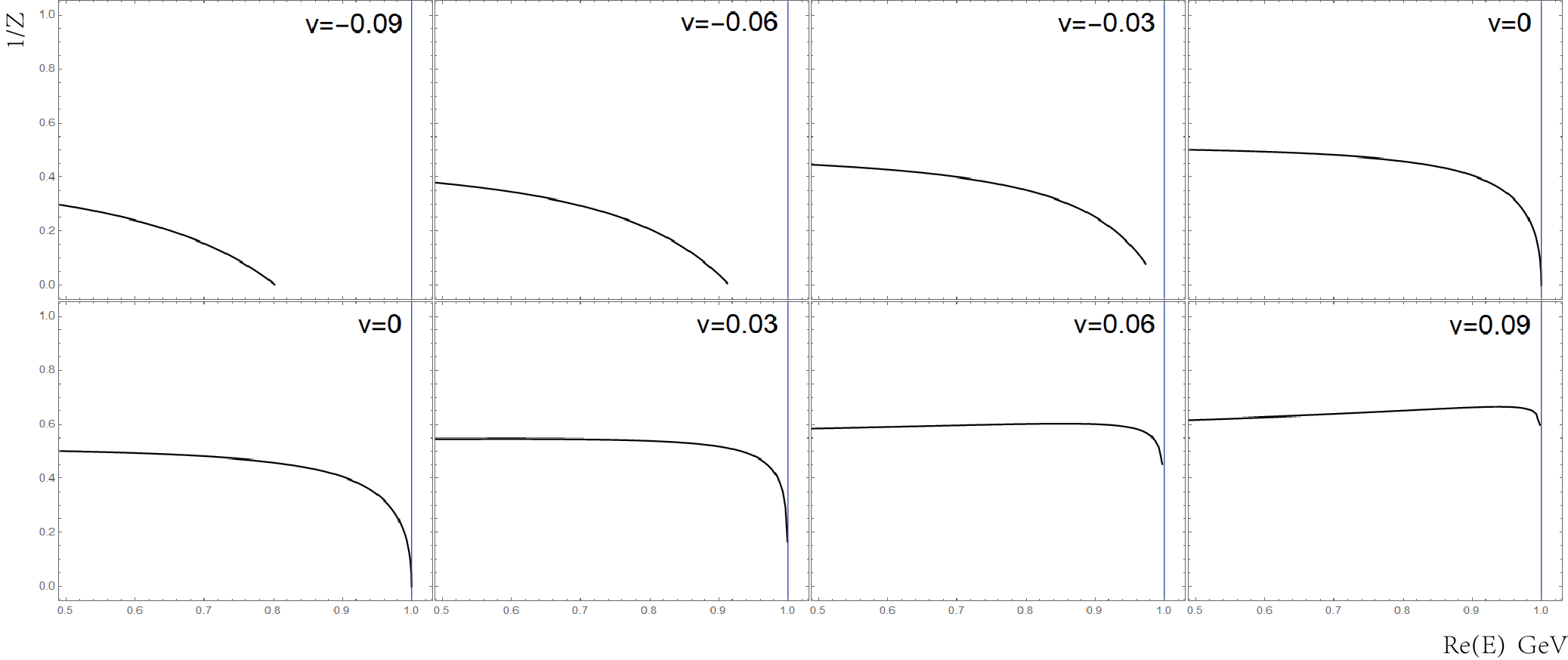}
    \caption{The $1/Z$ defined in Eq.~\ref{eq:zzzz} of the bound states in $S$-wave interaction when the bare state mass is greater than the threshold.}
    \label{fig:ps-large}
\end{figure*}

The $1/Z$ represented in Fig.~\ref{fig:ps-large} correspond to the cases of $S$-wave interaction and the bare state mass is larger than threshold. 
The results show that whatever the value of $v$, the proportion of bare state components will drop to zero when the binding energy is close to zero.
It indicates that for the $S$-wave bound state, when $v$ is attractive potential or even a weak repulsive potential, and the bare mass is above the threshold, the compositeness always become 1 when it is a shallow bound state.
For the $X(3872)$, it is just this case, thus the compositeness of $X(3872)$ is always close to 1, which conclusion is consistent with that in Ref.~\cite{Song:2022yvz}.

\begin{figure}
    \centering
    \includegraphics[width=0.8\linewidth]{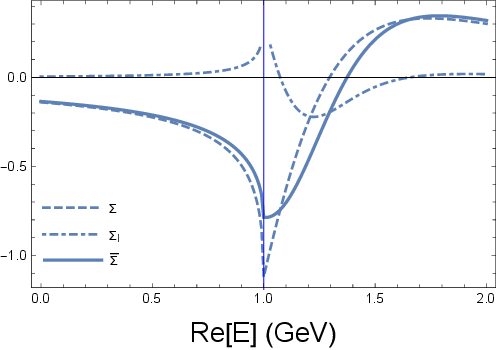}
    \caption{The variation law of the real part $\Sigma_{B}(E)$ [Dashed line], $\Sigma^I_{B}(E)$ [DotDashed line] and $\bar{\Sigma}(E)$ [Solid line] with binding energy $E$. Hear we set $g=0.3$, $v=0.01$. The vertical axis is the real part of the $\Sigma_{B}(E)$, $\Sigma^I_{B}(E)$ and $\bar{\Sigma}(E)$.}
    \label{fig:comp1}
\end{figure}

\begin{figure}
    \centering
    \includegraphics[width=0.8\linewidth]{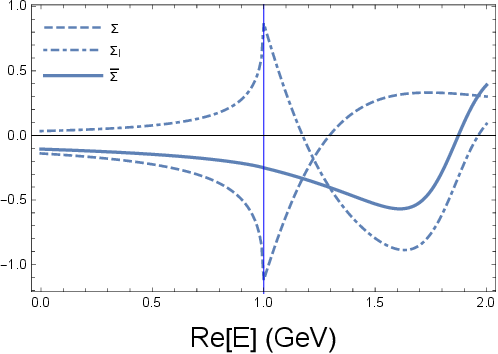}
    \caption{The variation law of the real part $\Sigma_{B}(E)$ [Dashed line], $\Sigma^I_{B}(E)$ [DotDashed line] and $\bar{\Sigma}(E)$ [Solid line] with binding energy $E$. Hear we set $g=0.3$, $v=0.09$. The vertical axis is the real part of the $\Sigma_{B}(E)$, $\Sigma^I_{B}(E)$ and $\bar{\Sigma}(E)$.}
    \label{fig:comp2}
\end{figure}

However, for the $v$ is a repulsive potential, it can be seen that the minimum $1/Z$ value at the threshold is not always 0. 
Especially when $v$ is large, as shown in the Fig.~\ref{fig:ps-large} with $v=0.06$ and $0.09$. 
By analyzing Eq.~(\ref{z}), we know that if we want $1/Z$ to be zero, the $\bar{\Sigma}(E)$ value at the threshold needs to be a cusp to ensure that its derivative is infinite. 
By plotting the distribution of Re$[\bar{\Sigma}(E)]$ at $v=0.01$ and $v=0.09$ as shown in Fig.~\ref{fig:comp1} and ~\ref{fig:comp2}.
The curves for Re$[\bar{\Sigma}(E)]$ show a clear cusp and a flat behavior at the threshold for $v=0.01$  and $v=0.09$, respectively, as shown the solid lines in Fig.~\ref{fig:comp1} and ~\ref{fig:comp2}.
This results in that $1/Z$ is very close to 0 for $v=0.01$, while a larger value for $1/Z$ instead of 0 for $v=0.09$. 

\begin{figure*}
    \centering
    \includegraphics[width=0.8\linewidth]{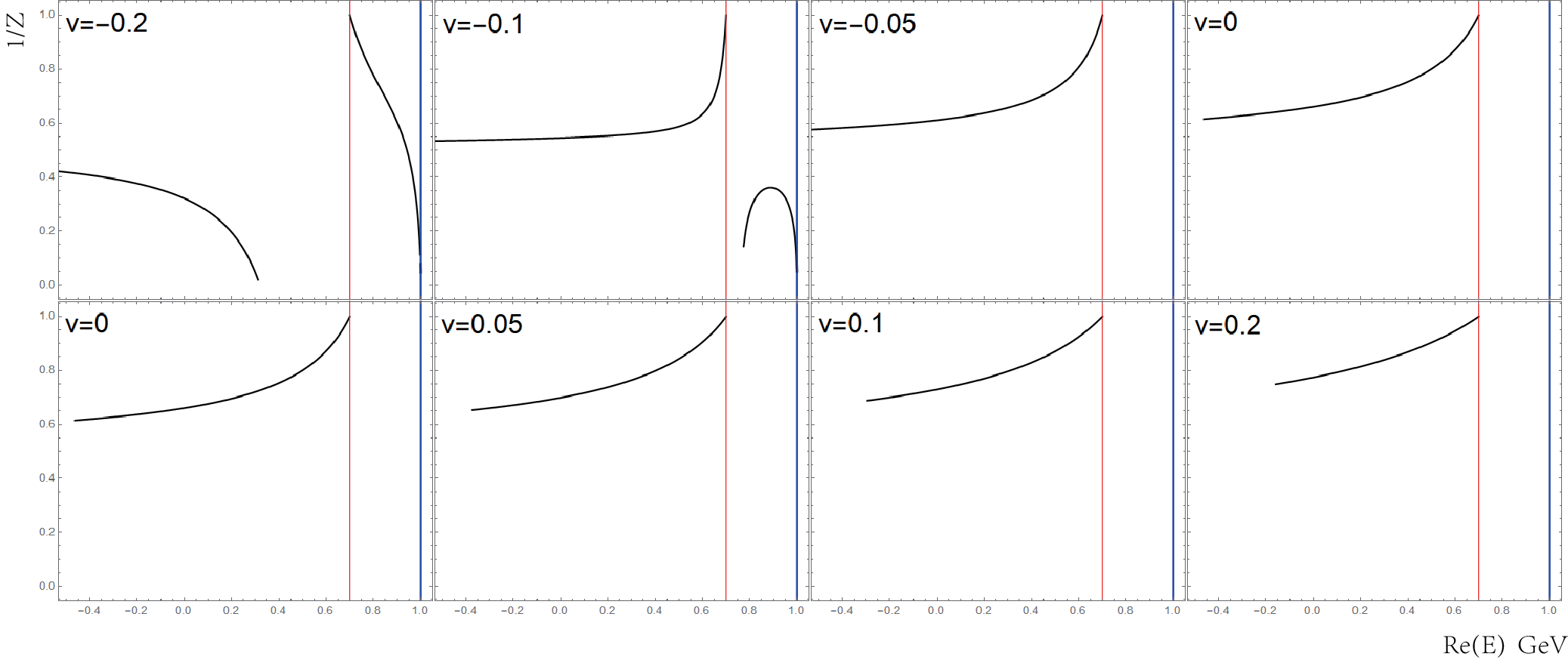}
    \caption{The $1/Z$ of the bound states in $S$-wave interaction when the bare state mass is smaller than the threshold.}
    \label{fig:ps-small}
\end{figure*}

Next, we considering the case that the bare state mass below threshold in $S$-wave. 
The results are shown in Fig.~\ref{fig:ps-small}. 
It shows interesting phenomenon that when bound state is close to the bare state, $1/Z$ will be close to 1 while it close to 0 when binding energy is close to 0.
Even for $v=-0.2$, the shallow bound state is originated from the bare state as shown in Fig.~\ref{fig:s-small}, but $1/Z$ is close to 0 for binding energy close to 0.
It indicates that even for the physical state which main components is coupled channel, i.e. compositeness close to 1, the bare state still play an important role, especially for the shallow bound state.

\begin{figure*}
    \centering
    \includegraphics[width=0.8\linewidth]{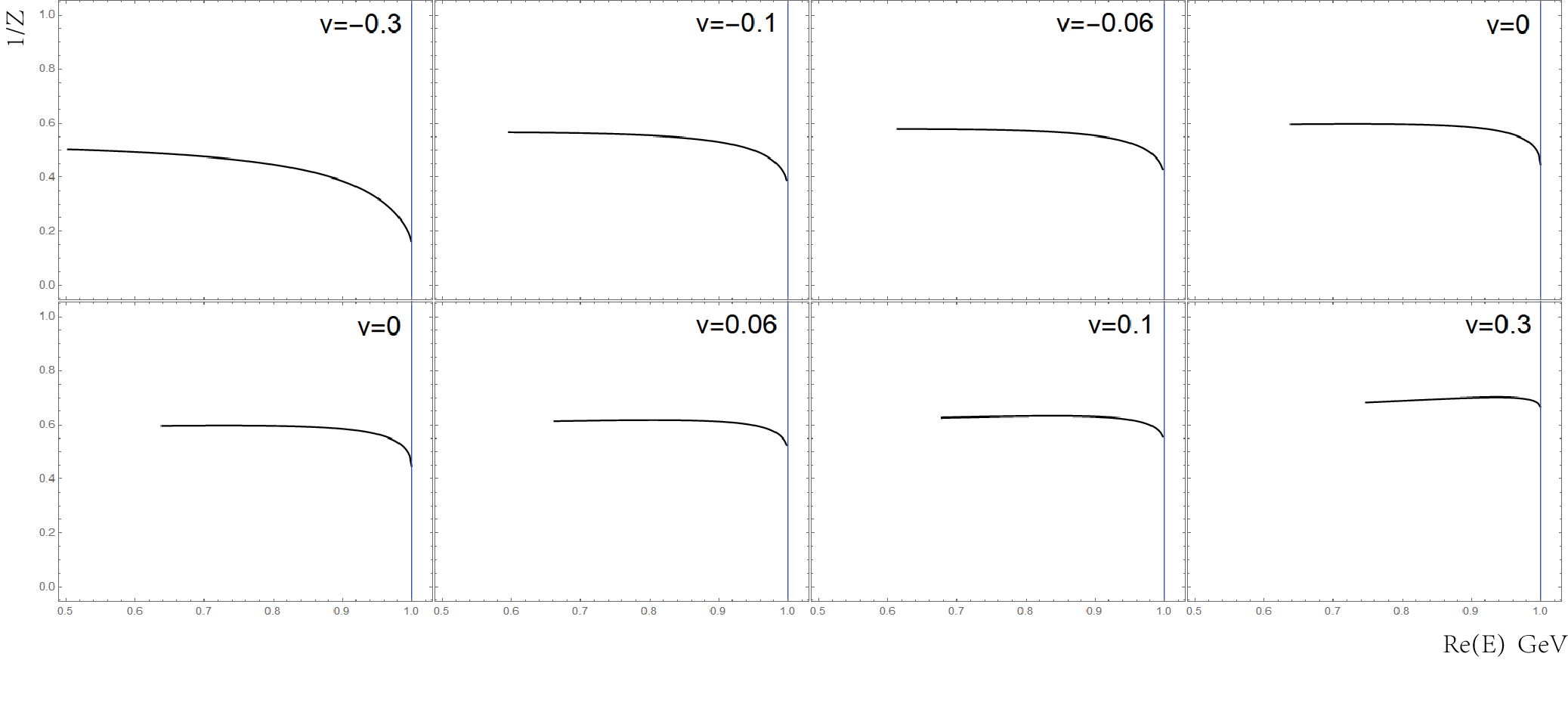}
    \caption{The $1/Z$ of the bound states in $P$-wave interaction when the bare state mass is greater than the threshold.}
    \label{fig:pp-large}
\end{figure*}

\begin{figure*}
    \centering
    \includegraphics[width=0.8\linewidth]{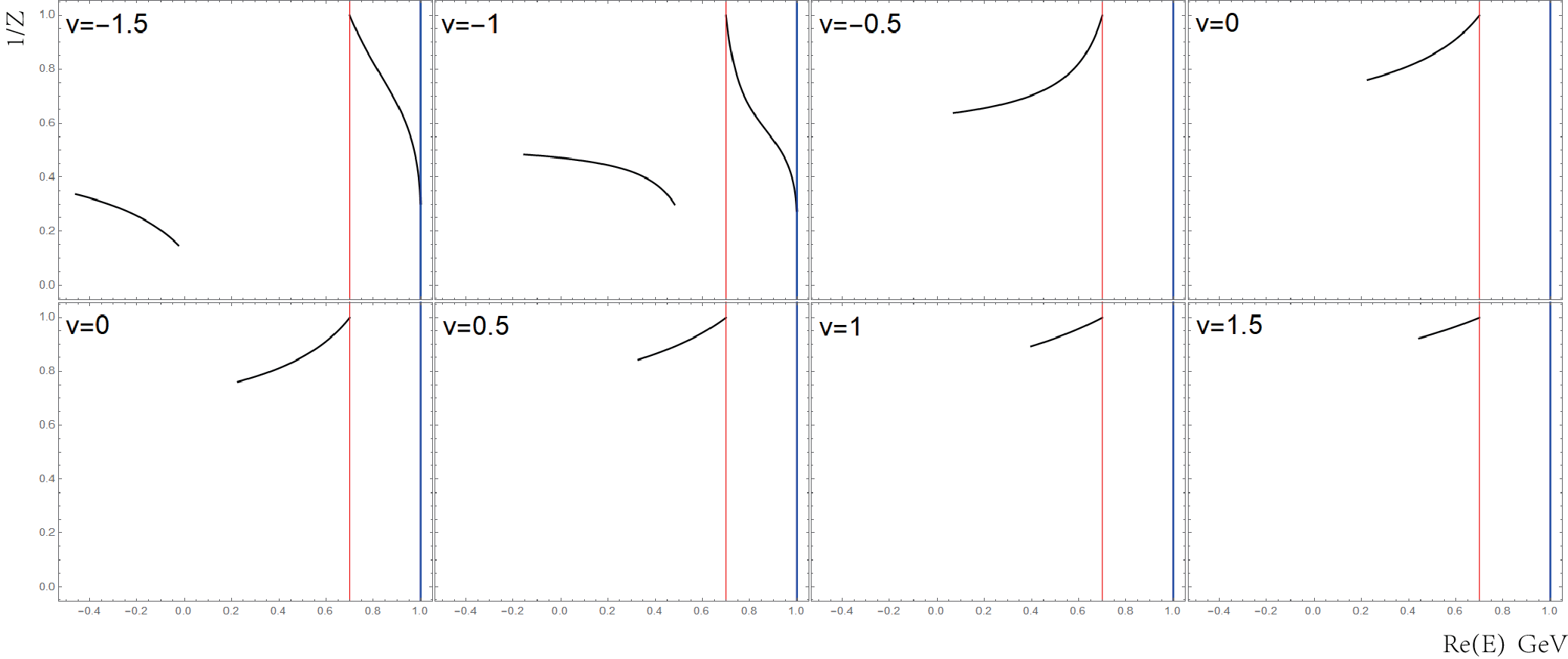}
    \caption{The $1/Z$ of the bound states in $P$-wave interaction when the bare state mass is smaller than the threshold..}
    \label{fig:pp-small}
\end{figure*}

At last, the $P$-wave cases with bare state mass larger and smaller than threshold are shown in Fig.~\ref{fig:pp-large} and \ref{fig:pp-small}, respectively.
In this case, even the bound state is close to the threshold, we find the compositeness will not to 1, since there is momentum dependence in the form factor in $f$ function which results in the absence of an extremum at the threshold.
Then, the compositeness ($1-1/Z$) will not be zero at the threshold as shown in Eq.~(\ref{z}).
Here we want to point out that for the $P$-wave and $S$-wave with the relativistic dispersion relationship, the loop function is ultraviolet divergence if there is no form factor to suppress the high momentum part.
Thus, the $\bar{\Sigma}(E)$ would suffer the uncertainties of the form factor which form actually is an assumption.
In this work, we just choose one form of form factor as a toy model to make the rough analysis, and to demonstrate the complexity of compositeness in a system containing a bare state and coupled channels. 
We suggest to make several different forms of form factor to confirm such uncertainties.

\section{Summary}
\label{sec:summary}

In this work, we investigate a system incorporating a bare state and coupled channels under the assumption of a separable potential. 
We solve the LS equation and systematically analyze the poles of the two-body scattering amplitude as well as the compositeness of the corresponding bound states, exploring their dependence on relevant parameters. 
Firstly, we qualitatively calculated the relationship between momentum and energy in two body scattering. 
We observed that compared with non-relativistic cases, there is a significant difference in the distribution of resonance state poles and bound state poles, and the distribution of energy momentum on different Riemann sheet is also different.
The complexity in relativistic situations arises from the existence of the square root in relativistic formulas by influencing the regional distribution of momentum sheet and energy sheet.

We then highlight that the form factor of the interaction influences the number of singularities in the scattering amplitude. 
Specifically, under the approximation of non-relativistic dispersion relations, the loop integral can be computed analytically. 
The higher the power of momentum in the denominator of the form factor, the higher the maximum degree of the polynomial in the energy function for the resulting loop integral, leading to multiple poles.
However, since general effective models are only valid near the threshold energy region, only a few of these poles are determined by the scattering amplitude near the threshold, while the others, located farther from the threshold, are strongly dependent on the specific form of the form factor. 
Consequently, the number of physically meaningful poles is significantly limited. 
Therefore, in the subsequent discussion, we primarily focus on the poles near the threshold.

On this basis, we investigated the properties of two-body scattering using the HEFT model and examined the impact of introducing a bare state on the two-body scattering states. 
By adjusting the coupling strength, we thoroughly studied the trajectories of the poles as they vary with the interaction strength.
Our results demonstrate that the presence of a bare state can influence the distribution of two-body scattering poles, thereby altering their properties. 
This indicates that when studying a two-body system, the influence of nearby coupled states cannot be simply ignored.

Furthermore, we calculated the proportion of the bare state component in the dressed state poles by analyzing the wave function, thereby quantifying the extent to which the bare state affects the bound state poles.
For the attractive interaction ($v<0$) between coupled channel and $S$-wave case, the results reveal that the closer the pole is to the threshold, the smaller the bare state component becomes, whereas the farther it is from the threshold, the larger the bare state component is.
However, for the repulsive potential, it is possible that the shallow bound state still has non-zero bare state components, which is related to the form factor.

\section{Acknowledgements}

We thank helpful discussions with Lei Chang, Feng-kun Guo, and Bing-Song Zou.
This work is supported by the National Natural Science Foundation of China under Grant Nos. 12175239 and 12221005, 
and by the Chinese Academy of Sciences under Grant No. YSBR-101,
and by the National Key R$\&$D Program of China 2024YFA1610502.

\bibliographystyle{unsrt}%规定参考文献的样式
\bibliography{cite}  %参考文献库的名字Ref

\end{document}